quant-ph/0602200

\documentclass[prl,twocolumn,showpacs]{revtex4}
\usepackage{graphicx}

\begin{document}

 \def\BE{\begin{equation}}
 \def\EE{\end{equation}}
 \def\BEA{\begin{eqnarray}}
 \def\EEA{\end{eqnarray}}
 \def\BA{\begin{array}}
 \def\EA{\end{array}}
 \def\r{{\vec\rho}}
 \def\q{{\vec q}}
 \def\O{\Omega}

 \title{Quantum teleportation of optical images with frequency conversion}
 \author{Liubov V.~Magdenko, Ivan V.~Sokolov}
 \affiliation{V.~A.~Fock Physics Institute, St.-Petersburg University,
       198504 Stary Petershof, St.-Petersburg, Russia}
 \author{Mikhail I.~Kolobov}
 \affiliation{Laboratoire PhLAM, Universit\'e de Lille-1,
       F-59655 Villeneuve d'Ascq cedex, France}

\begin{abstract}
We describe a new version of continuous variables quantum
holographic teleportation of optical images. Unlike the previously
proposed \cite{Sokolov01, Gatti04} scheme, it is based on the
continuous variables quantum entanglement between the light fields
of different frequencies and allows for the wavelength conversion
between the original and the teleported images. The frequency
tunable holographic teleportation protocol can be used as a part
of light-matter interface in quantum {\it parallel} information
processing and {\it parallel quantum memory}.
\end{abstract}

%\ocis{270.6570, 100.6640}
%\pacs{PACS number(s): 42.50.Dv, 42.30.Wb, 42.50.Lc}
\pacs{42.50.Dv, 42.30.Wb, 42.50.Lc}
%\date{\today}
 \maketitle

Continuous variables quantum teleportation allows to transfer an
arbitrary quantum state of the electromagnetic field between two
spatially separated systems via an exchange of classical
information in combination with quantum entanglement shared by
these systems. Initially the protocol of continuous variable
teleportation was proposed
theoretically~\cite{Vaidman94,Braunstein98a} and realized
experimentally\cite{Furusawa98,Bowen03} for a single spatial mode
of the electromagnetic field. Recently this protocol was
generalized theoretically to spatially multimode electromagnetic
fields~\cite{Sokolov01,Gatti04}. The important feature of
spatially multimode teleportation scheme is that it allows for
simultaneous parallel teleportation of optical images (still or
time-varying) containing large number of elements or pixels. The
proposed scheme was called {\it holographic teleportation} because
it resembles the conventional holography with an important
difference that the reconstructed image is a quantum copy of the
original one with fidelity that can be made close to unity. It is
impossible to achieve high fidelity in holographic teleportation
without sharing multimode quantum entanglement by the original and
the target systems.

One of possible applications of quantum teleportation is the
light-matter interface for quantum memory\cite{Polzik04} for light
which allows to record and story a quantum state of light on that
of an atomic ensemble. The generalized spatially multimode
teleportation scheme opens new perspectives for creation of a {\it
parallel quantum memory} for parallel processing of quantum
information. For an efficient interaction between the light wave
and the atomic medium it is desirable to have a possibility for
tuning the optical frequency of light to that of the atomic
transition without changing the quantum state of the
electromagnetic field. Let us mention a proposal of quantum
frequency conversion\cite{Kumar92} that was the first
demonstration of quantum state transfer into different frequency.
In recent experiment\cite{Tanzilli05} quantum transfer of qubits
was successfully demonstrated between photons of wavelength 1310
nm and 710 nm.

In this letter we describe a new version of holographic
teleportation of optical images with {\it frequency conversion}.
This means that an input image, carried by a light wave with
optical frequency $\omega_1$, is teleported to the output image at
different optical frequency $\omega_2$ with preservation of its
original quantum state. Similar to the frequency-preserving
quantum holographic teleportation, discussed in
Refs.~\cite{Sokolov01, Gatti04}, to obtain high fidelity in the
teleported image one needs to create spatially multimode
entanglement between the original and the target systems. In our
case the source of spatially multimode entanglement between the
input image at frequency $\omega_1$ and the teleported image at
frequency $\omega_2$ is a type-I traveling-wave non-degenerate
optical parametric amplifier (OPA).

The optical scheme of holographic teleportation with frequency
conversion if shown in Fig.~1. An input image which is to be
teleported from Alice to Bob, is described by a slowly-varying
field operator ${\hat A}^{\rm in}_1(\r,t)$, where $\r=(x,y)$ is a
two-dimensional transverse coordinate. An input electromagnetic
wave carrying this image is splitted into two secondary waves by a
50/50 beam splitter BS$_1$, and two quadrature components of these
secondary waves are homodyne detected by two difference detectors
using two local oscillators LO$_x$ and LO$_y$, two 50/50 beam
splitters BS$_2$ and BS$_3$, and four efficient CCD cameras with
appropriate spatial resolution.
 \begin{figure}
 \includegraphics[width=7cm]{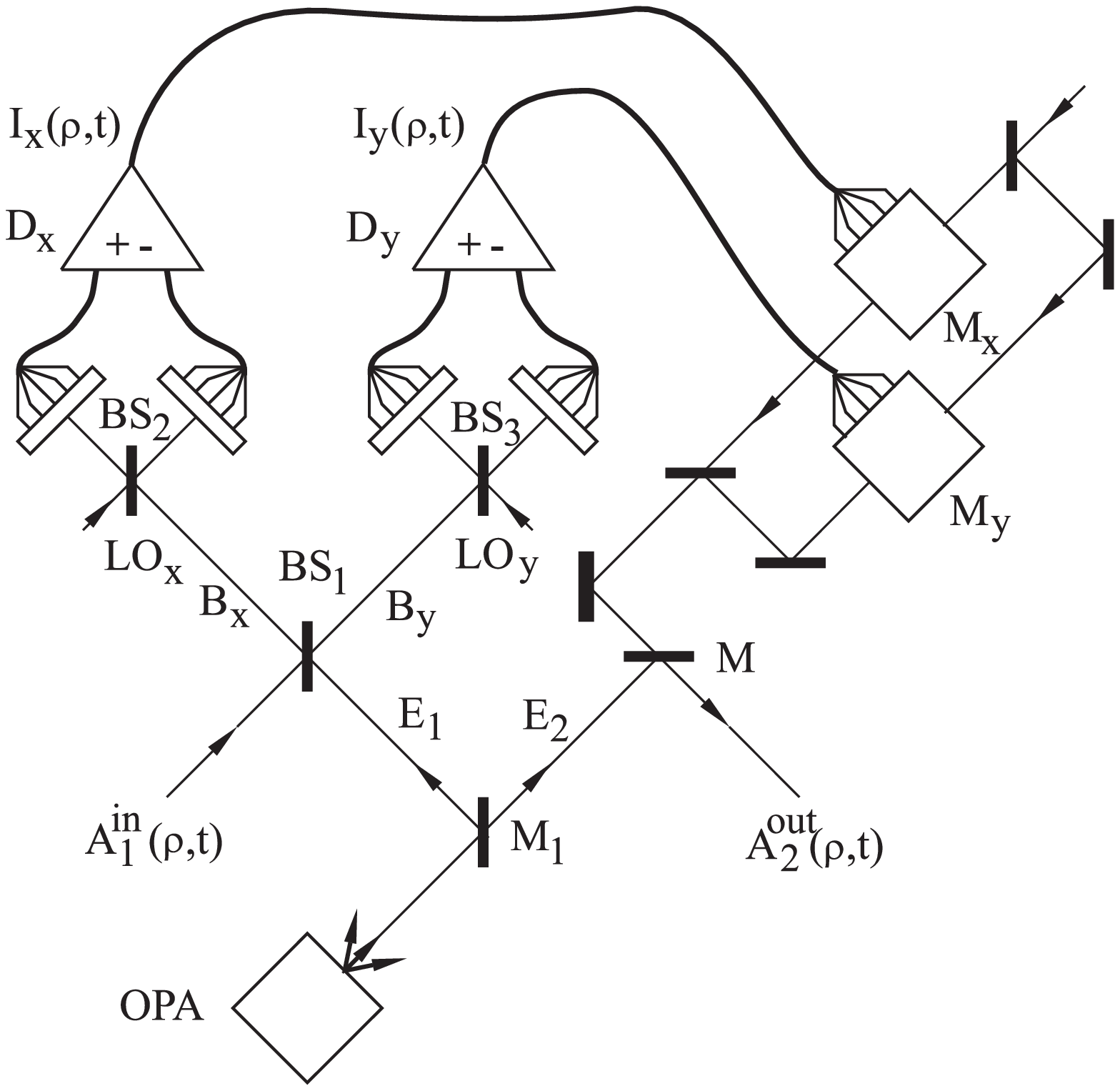}
 \caption{Optical scheme of holographic teleportation with frequency
conversion.}
 \end{figure}
The local oscillator waves have the same frequency $\omega_1$ as
the carrier frequency of the input image. The slow-varying
difference photocurrent densities from these CCD cameras,
containing information about spatio-temporal quantum fluctuations
of the input image, are transmitted from Alice to Bob and are used
for preparation of the output field ${\hat A}_2(\r,t)$. The
multi-channel modulators M$_x$ and M$_y$ perform spatio-temporal
modulation of a coherent wave with different carrier frequency
$\omega_2$. This part of the scheme can be viewed as a classic
non-stationary holography, where the reconstructing wave has
different wavelength.

A key ingredient of our teleportation scheme is a pair of
spatially-multimode EPR fields ${\hat E}_n(\r,t)$, $n=1,2$. Since
an entanglement between different carrier frequencies is needed,
the fields are created by a type-I traveling-wave {\sl
non-degenerate} OPA. The dichroic mirror M$_1$ reflects the wave
at frequency $\omega_1$ and transmits that at frequency
$\omega_2$. In the case of perfect reflectivity there are no
frequency matched vacuum fluctuations entering from the open ports
of M$_1$ into the corresponding fields.

Let us introduce slowly-varying spatio-temporal annihilation and
creation operators ${\hat E}_n(\r,t)$ and ${\hat
E}^{\dag}_n(\r,t)$, $n=1,2$ of the electromagnetic waves with
central frequencies $\omega_1$ and $\omega_2$ at the output of the
non-degenerate OPA. The frequencies $\omega_1$ and $\omega_2$ obey
the condition of energy conservation, $\omega_1+\omega_2 =
\omega_p$, where $\omega_p$ is the frequency of the pump wave. The
field operators are normalized so that $\langle {\hat
E}_n^{\dag}({\r},t) {\hat E}_n({\r},t) \rangle$ gives the mean
value of the irradiance, expressed in photons per ${\rm cm}^2$ per
second.

The transformation of the input fields ${\hat A}_n(\r,t)$ of the
non-degenerate OPA in the vacuum state into the output fields ${\hat
E}_n(\r,t)$ in the broadband multimode squeezed state is described in
terms of the Fourier components of these operators in frequency and
spatial-frequency domain, ${\hat E}(\r,t) \rightarrow {\hat
e}(\q,\Omega)$. The squeezing transformation performed by a non-degenerate
OPA, can be written as follows:
\BE
      {\hat e}_n(\q,\Omega) = U_n(\q,\Omega) {\hat a}_n(\q,\Omega) +
      V_n(\q,\Omega) {\hat a}_{n'}^{\dag}(-\q,-\Omega),
                      \label{squeezing}
\EE
where $n=1,2$,  $n\neq n'$, the coefficients $U_n(\q,\Omega)$ and
$V_n(\q,\Omega)$ depend on the amplitude of the pump field,
nonlinear susceptibility and the phase-matching condition. The
explicit form of $U_n(\q,\Omega)$ and $V_n(\q,\Omega)$ can be
found, for example, in Ref.~\cite{Brambilla04}. The EPR
correlations between the fields ${\hat E}_1(\r,t)$ and ${\hat
E}_2(\r,t)$ are determined by two parameters, namely, the
orientation angle $\psi_n(\q,\Omega)$ of the major axis of the
squeezing ellipse~\cite{Kolobov99} and the degree of squeezing
$r_n(\q,\Omega)$,
 \BE
     \psi_n(\q,\Omega) = \frac{1}{2}\arg\left\{U_n(\q,\Omega)
     V_{n'}(-\q,-\Omega)\right\},
                    \label{psi}
 \EE
 \BE
    \exp\left[\pm r_n(\q,\Omega)\right] = |U_n(\q,\Omega)|
    \pm |V_{n'} (\q,\Omega)|.
                    \label{exp_r}
 \EE
It should be noted that in Fig.~1 we have chosen the input field
at the frequency $\omega_1$ and the output field at the frequency
$\omega_2$, that corresponds to $n=1$ and $n'=2$ in
Eqs.~(\ref{psi}), (\ref{exp_r}), but with the same source of
entangled beams the teleportation with frequency conversion can be
performed from $\omega_2$ to $\omega_1$ as well.

It is instructive for better understanding of the scheme to give a simple
physical explanation for the continuous variables quantum entanglement
arising from squeezing in {\sl non-degenerate} parametric down-conversion.
For simplicity we shall omit the spatial dependence of the fields. Let us
introduce the sum of two slow amplitudes, ${\cal E}(t) \sim E_1(t)+
E_2(t)$ and consider a contribution to ${\cal E}(t)$ from a pair of
Fourier amplitudes,
 \BE
{\cal E}(t) \sim e_1(-\Omega) e^{i\Omega t} + e_2(\Omega)
e^{-i\Omega t}.
 \label{pair_amplitudes}
 \EE
One can show~\cite{Magdenko06}, that the discussed above squeezing ellipse
for given $\Omega$ represents exactly the quantum uncertainty region for
this contribution. For perfect squeezing this automatically implies a
perfect phase matching beyond the classical limit between the slow
amplitudes in the right side of (\ref{pair_amplitudes}), as illustrated in
Fig.~2a. Let us remind that these amplitudes were defined with respect to
the different carrier frequencies $\omega_1$ and $\omega_2$. Hence the
corresponding beams can be split by a dichroic mirror, preserving their
quantum entanglement.
 \begin{figure}
 \includegraphics[width=7.5cm]{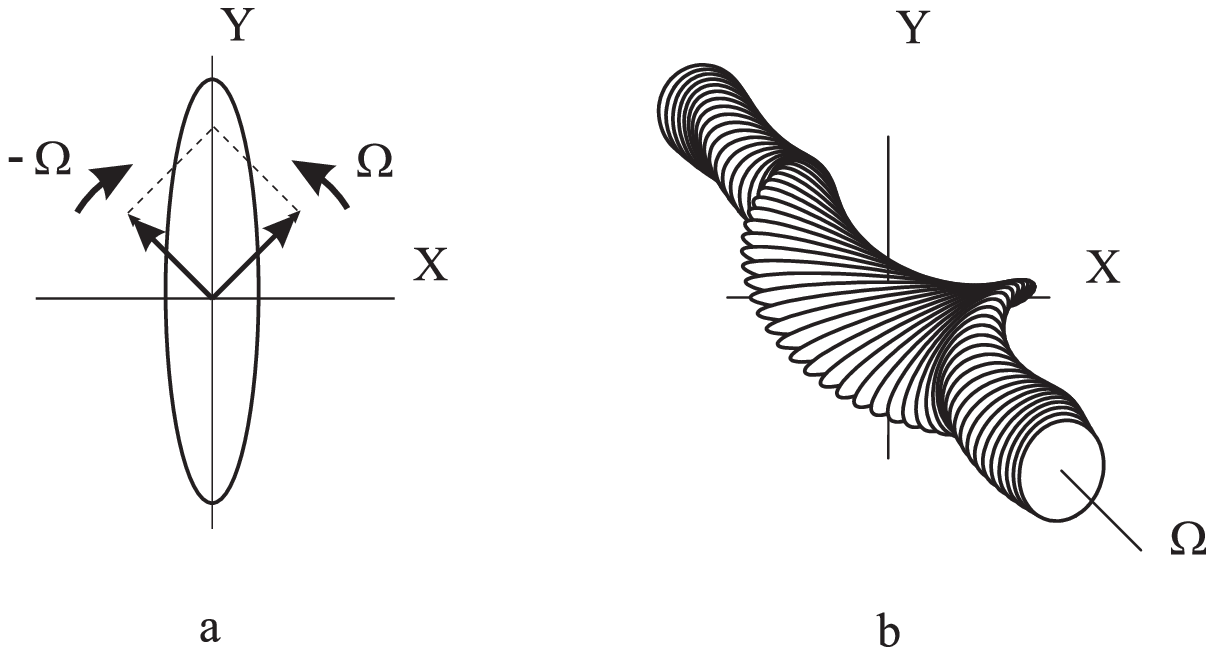}
 \caption{ Uncertainty region for the slow field amplitude from
Eq.~4 (a), the frequency dispersion of squeezing ellipses for
non-degenerate traveling-wave OPA, where $r_2(0,0)=3$, $\Omega$ --
arb.~units (b). Here X, Y -- field quadrature components.}
 \end{figure}
The first slow amplitude is imprinted into the photocurrents by the
homodyne detection and  into the reconstructing field at different
frequency $\omega_2$ by the modulation. The second one {\sl ab initio} is
related to this frequency. A proper matching of the relative amplitudes
and phases in quantum and classical channels leads to the quantum noise
cancellation and effective teleportation.

Next important point is that the quantum modulation of the output field at
a given frequency is now due to two independent pairs of entangled
amplitudes, the first pair is shown in (\ref{pair_amplitudes}), another
one corresponds to $\Omega \rightarrow -\Omega$. Therefore, in order to
achieve optimal results one has to take care of a proper phase matching of
two independent squeezing ellipses (at $\Omega$ and $-\Omega$). The
frequency dependence of squeezing within the frequency range of effective
non-degenerate parametric down-conversion is graphically shown in Fig.~2b.

Apart from these new features, introduced by the frequency
non-degenerate nature of the process, the fields evolution in the
scheme is very similar to that described in Refs.~\cite{Sokolov01,
Gatti04}. The input field ${\hat A}^{\rm in}_1(\r,t)$ is mixed
with one EPR beam ${\hat E}_1(\r,t)$ at the 50/50 beam splitter
BS$_1$. The secondary waves ${\hat B}_x$ and ${\hat B}_y$,
 \BE
    {\hat B}_{x,y}(\r,t) = \frac{1}{\sqrt{2}}\big(\pm {\hat A}^{\rm in}_1(\r,t) +
    {\hat E}_1(\r,t)\big),
 \EE
with the $+(-)$ sign corresponding to $x(y)$ channel, are
photodetected by means of two balanced homodyne detectors. The
difference photocurrents $I_x(\r,t)$ and $I_y(\r,t)$ collected
from individual pixels of these matrices, carry the information
about the spatio-temporal quantum fluctuations of the quadrature
components of ${\hat B}_x(\r,t)$ and ${\hat B}_y(\r,t)$, phase
matched with LO$_x$ and LO$_y$. These photocurrents are sent from
Alice to Bob via two multichannel classical communication lines
and are used by Bob for spatio-temporal modulation of an external
coherent wave with the frequency $\omega_2$, by means of two
multichannel modulators $M_x$ and $M_y$. The teleported field
${\hat A}^{\rm out}_2(\r,t)$ is created by mixing of this
modulated wave with a second EPR wave ${\hat E}_2(\r,t)$ at the
mirror M with high reflectivity. By choosing appropriately the
mirror transmission and the modulation depth, the teleported field
${\hat A}^{\rm out}_2(\r,t)$ can be obtained as
 \BE
       {\hat A}^{\rm out}_2(\r,t) = {\hat A}^{\rm in}_1(\r,t)+{\hat
       F}(\r,t),
                        \label{field_out}
 \EE
where ${\hat F}(\r,t)$ is an operator describing the noise added
to the teleported field,
 \BE
  {\hat F}(\r,t) = {\hat E}_2(\r,t) +
       {\hat E}_1^{\dag}(\r,t).
                        \label{noise}
 \EE
It can be demonstrated that the added noise can be considered as
classical, thus preserving the commutation relations for the
output field.

An ideal teleportation from $\omega_1$ to $\omega_2$ would
correspond to the situation when the output field operator at
different spatial points $\r$ and at different time moments $t$ is
an exact copy of the input field operator, ${\hat A}^{\rm
out}_2(\r,t) = {\hat A}^{\rm in}_1(\r,t)$. However, as explained
in Refs.~\cite{Sokolov01, Gatti04}, this would require an infinite
energy of the EPR beams and therefore could never be realized in
practice. In real experimental situation the teleportation process
will never take place ``point-to-point'', but on average within
some finite spatial area and within some finite time interval. In
order to quantitatively describe the performance of our
teleportation scheme we  introduce a coarse-grained description of
the input and output fields. Precisely, we consider the averaged
field operator over a square pixel $S_j$ of area $S=\Delta^2$ and
over a time interval $T_i$ of duration $T$:
 \BE
     {\hat A}^{\rm out}_2(j,i)= \frac{1}{\sqrt{ST}}\int_{S_j}d\r
     \int_{T_i} dt\, {\hat A}^{\rm out}_2(\r,t),
                      \label{averaged_field}
 \EE
with analogous definitions for the input field. The averaged field
operators obey standard commutation relations of discrete field
oscillators. The quadrature components of the averaged field
operators at the output are
 \BEA
       {\hat X}^{\rm out}_{\varphi}(j,i) &=& {\hat A}_2^{\rm out}(j,i) e^{-i\varphi}
       + h.\,c., \nonumber \\
       {\hat Y}^{\rm out}_{\varphi}(j,i) &=& -i {\hat A}_2^{\rm out}(j,i)
       e^{-i\varphi}
       + h.\,c.,
                     \label{averaged_quadratures}
 \EEA
and similarly for the input field operators.

To characterize the noise added in the teleportation process we
compare the correlation functions of the input and output
quadrature components defined in Eq.~(\ref{averaged_quadratures}).
As follows from Eq.~(\ref{field_out}), the relation between these
correlation functions is
 \BE
    \langle \delta {\hat X}_\varphi^{\rm out} (j,i)\,\delta
    {\hat X}_\varphi^{\rm out}(j',i')\rangle =
                       \label{correlation}
 \EE
 $$
 \langle \delta {\hat X}_\varphi^{\rm in} (j,i)\,
    \delta {\hat X}_\varphi^{\rm in} (j',i') \rangle
    + {\cal C} (j,j'; i,i'),
 $$
where ${\cal C} (j,j'; i,i')$ is the covariance matrix of the added noise.
The explicit expression of this covariance matrix is
 \BE
   {\cal C}  \left( j,j' \,;\, i, i'\right)
   =  2  \int {\rm d}\q \int {\rm d}\Omega \, B_\Delta (\q\,) B_T
   (\Omega)\times
                          \label{covariance}
 \EE
 $$
 \cos{ \left[ \q\cdot(\r_{j} - \r_{j'}) -\Omega (t_i -t_{i'}) \right] }
   G(\q,\Omega),
 $$
where $\r_j=\{x_j,y_j\}$ is the center of the j-th pixel, and $t_i$ is the
center of the i-th time interval. The functions $B_\Delta(\q\,)$ and
$B_T(\Omega)$ arise from the coarse-graining operation, and for a square
pixel of size $\Delta$ they read,
 \BEA
    B_\Delta(\q\,) &=& \frac{\Delta^2}{4\pi^2}{\rm sinc^2}
    \left(\frac{q_x \Delta}{2} \right)
    {\rm sinc^2} \left(\frac{q_y \Delta}{2} \right),\nonumber \\
    B_T(\O) &=& \frac{T}{2\pi}{\rm sinc^2}
    \left(\frac{\O T}{2} \right).
                \label{B_delta_T}
 \EEA
The Green function $G(\q,\Omega)$ expressed in terms of the
orientation angle $\psi_2(\q,\Omega)$ and the degree of squeezing
$r_2(\q,\Omega)$ from Eq.~(\ref{psi}) looks like
 \BE
     G(\q,\Omega) = e^{2r_2(\q,\Omega)}
     \cos^2\psi_2(\q,\Omega) + e^{-2r_2(\q,\Omega)}
     \sin^2\psi_2(\q,\Omega).
                  \label{green}
 \EE
As follows from Eqs.~(\ref{covariance})-(\ref{green}), the covariance
matrix ${\cal C} (j,j'; i,i')$ is independent of the phase $\varphi$ in
Eq.~(\ref{averaged_quadratures}). Thus, the added noise is the same for
any quadrature.

In the absence of the EPR correlations, i.~e.~when
$r_2(\q,\Omega)=0$, we obtain the classical limit of teleportation
with the covariance matrix ${\cal C}_{\rm cl} (j,j'; i,i') =
2\delta_{jj'}\delta_{ii'}.$ In this limit two units of vacuum
noise are added at each pixel exactly as for a single-mode
teleportation and for the teleportation of images without
frequency conversion.

Choosing the phase matching condition in the OPA such that
$\psi_2(0,0)=\pi/2$, we obtain reduction of the quantum noise
below the standard quantum level within some bandwidth of spatial
frequencies $\q$ and temporal frequencies $\Omega$ determined by
the phase-matching conditions in the crystal.  When the pixel size
$\Delta$ and the time window $T$ are much larger than the
characteristic coherence length $l_c$ and the coherence time $T_c$
of the OPA, we obtain
 \BE
    \lim_{\Delta \to \infty \; , T \to \infty}
    {\cal C}_{\rm cl} (j,j'; i,i')=2\delta_{jj'}\delta_{ii'}
    \exp[-2r(0,0)].
                \label{quant_limit}
 \EE
In Fig.~3 we illustrate the role of the pixel size $\Delta$ and of the
integration time $T$ in the teleportation process with frequency
conversion. This figure shows the diagonal elements ${\cal C} (j,j; i,i)$
of the covariance matrix as a function of the relative pixel size
$D=\Delta/l_c$ for three different observation times $T$ equal to $10T_c$,
$T_c$, and $0.1T_c$. The coherence time $T_c$ for a frequency
non-degenerate OPA is typically estimated as the time delay at the crystal
length $l$ between two wave packets centered at the frequencies $\omega_1$
and $\omega_2$, $T_c \sim l |1/v_1 - 1/v_2|$, arising due to the
difference of the group velocities $v_n$. Since a typical frequency
spectrum of parametric down-conversion is fairly broad, a wave-packet
spread due to the group velocity dispersion can also have an effect on
squeezing and entanglement in our teleportation scheme, as we shall
discuss elsewhere~\cite{Magdenko06}.
 \begin{figure}
 \includegraphics[width=7cm]{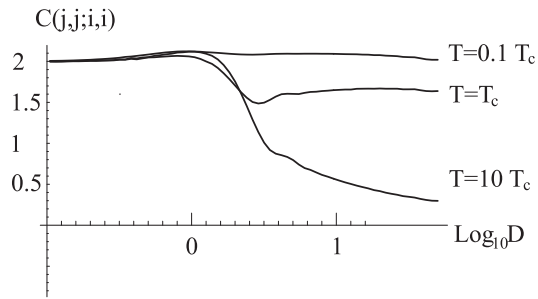}
 \caption{Diagonal elements ${\cal C} (j,j; i,i)$ of the noise
covariance matrix  as a function of the relative pixel size
$D=\Delta/l_c$ for three different observation times $T$ equal to
$10T_c$, $T_c$, and $0.1T_c$, here $r_2(0,0)=3$.}
 \end{figure}

The frequency dependence of squeezing orientation angle, arising in a
parametric crystal and illustrated in Fig.~2b, is a pure phase effect and
can be compensated by propagation in a linear medium with a properly
chosen frequency dependence of the refraction index. It is worth noting
that in many interference experiments with twin photons, a need for
similar compensation has been realized some time ago. Such a compensation
can be applied to the plots shown in Fig.~3 and will be illustrated in the
forthcoming publication~\cite{Magdenko06}.

The coherence time for a non-degenerate OPA is usually much larger
than for a degenerate one. Therefore, the assumption $T\gg T_c$
used in Refs.~\cite{Sokolov01, Gatti04} is not necessarily true
anymore. The role of the observation time $T$ is as crucial in the
teleportation process as the role of the pixel size $\Delta$.
Indeed, as follows from Fig.~3, the diagonal elements of the
covariance matrix which characterize the added noise power,
decrease with growing pixel size $\Delta$ from the classical limit
${\cal C} (j,j; i,i)=2$ to the EPR limit ${\cal C} (j,j;
i,i)=2\exp[-2r(0,0)]$ when the observation time is large compared
with the coherence time $T_c$. However, when the observation time
$T$ becomes comparable with the coherence time $T_c$ or smaller,
the diagonal elements ${\cal C} (j,j; i,i)$ never reach the EPR
limit but remain close to the classical limit even for large pixel
size.

This results allow to estimate an effective number of
spatio-temporal degrees of freedom for an input non-stationary
image that can be teleported in our scheme.

In conclusion we have proposed a new scheme of quantum holographic
teleportation of optical images from one optical frequency to
another. We have presented preliminary analysis of the performance
of our scheme and have revealed the possibilities to achieve high
quality teleportation. More detailed calculations including
evaluation of multi-pixel fidelity for this scheme are in progress
and will be published elsewhere. This work was supported by the
Network QUANTIM (IST-2000-26019) of the European Union and by the
INTAS under Project No.~2001-2097. The research was performed
within the framework of GDRE "Lasers et techniques optiques de
l'information".

\end{document}